\documentclass[letterpaper, 10 pt, conference]{ieeeconf}  

\IEEEoverridecommandlockouts                              
\usepackage{graphicx}
\usepackage{float}
\usepackage{booktabs}
\usepackage{array}
\usepackage{amsmath} 
\usepackage{url}
\usepackage{setspace}
\usepackage{tikz}
\usepackage{lipsum}
\usepackage{comment}
\usepackage{amssymb}
\usepackage{multirow}
\usepackage{cite}
\usepackage{tabularx}
\usepackage{algpseudocode}
\usepackage{subfig}
\usepackage{soul}
\usepackage[ruled,vlined]{algorithm2e}
\usepackage{siunitx}
\usepackage{mathtools}

\DeclareMathOperator*{\argmin}{arg\,min}

\title{\LARGE \bf
Energy-optimal Three-dimensional Path-following Control of Autonomous Underwater Vehicles under Ocean Currents}

\author{Niankai Yang$^{1}$, Chao Shen$^{2}$, Matthew Johnson-Roberson$^{3}$, and Jing Sun$^{1}$  
\thanks{$^{1}$Niankai Yang and Jing Sun are with the Department of Naval Architecture and Marine Engineering, University of Michigan, Ann Arbor, MI 48109, USA (\{ynk,\;jingsun\}@umich.edu).}
\thanks{$^{2}$Chao shen is with the Department of Systems and Computer Engineering, Carleton University, Ottawa, ON, Canada. (e-mail: shenchao@sce.carleton.ca).}
\thanks{$^{3}$Matthew Johnson-Roberson was with Department of Naval Architecture and Marine Engineering, University of Michigan, Ann Arbor, MI 48109, USA. He is now with the Robotics Institute in the School of Computer Science, Carnegie Mellon University, Pittsburgh, PA 15213, USA. (e-mail: mkj@andrew.cmu.edu).}
}

\begin{document}

\maketitle

\newcommand{\tabincell}[2]{\begin{tabular}{@{}#1@{}}#2\end{tabular}}

\begin{abstract}

This paper presents a three-dimensional (3D) energy-optimal path-following control design for autonomous underwater vehicles subject to ocean currents. The proposed approach has a two-stage control architecture consisting of the setpoint computation and the setpoint tracking. In the first stage, the surge velocity, heave velocity, and pitch angle setpoints are optimized by minimizing the required vehicle propulsion energy under currents, and the line-of-sight (LOS) guidance law is used to generate the yaw angle setpoint that ensures path following. In the second stage, two model predictive controllers are designed to control the vehicle motion in the horizontal and vertical planes by tracking the optimal setpoints. The proposed controller is compared with a conventional LOS-based control that maintains zero heave velocity relative to the current (i.e., relative heave velocity) and derives pitch angle setpoint using LOS guidance to reach the desired depth. Through simulations, we show that the proposed approach can achieve more than $13\%$ energy saving on a lawnmower-type and an inspection mission under different ocean current conditions. The simulation results demonstrate that allowing motions with non-zero relative heave velocity improves energy efficiency in 3D path-following applications.

\end{abstract}


\section{Introduction}

Autonomous underwater vehicles (AUV) are essential in achieving various underwater missions (e.g., seafloor mapping~\cite{zhang2013general} and underwater structure inspection~\cite{kondo2004navigation}). It has been projected that the market size of AUVs will be doubled in the next five years~\cite{AUVreport2020}. Particularly, demands for scientific and commercial uses of AUVs are expected to increase significantly, calling for more reliable and energy-efficient AUV platforms\cite{palomer2019inspection}. To enhance the reliability and the energy efficiency of an AUV, a well-designed control system is of great importance. 

Given that underwater missions typically involve following piecewise linear paths to cover a designated area, studies have been devoted to designing path-following control strategies for reliable AUV operations. For example, backstepping techniques~\cite{lapierre2007nonlinear} and model predictive control (MPC)~\cite{shen2018path} have been applied for two-dimensional (2D) path following of AUVs. The line-of-sight (LOS) guidance widely used in aerospace and surface vessel applications was adopted in~\cite{caharija2012relative} for the 2D AUV path following under ocean currents, and an extension to the 3D path following using LOS guidance was presented in~\cite{lekkas2013line}. Although the above approaches can complete the mission reliably, vehicle energy efficiency is not explicitly considered.

To account for energy efficiency in control designs, optimization has been incorporated to compute vehicle thrusts by minimizing a cost function, including control efforts. For instance, a weighted sum of reference-following error and control inputs was used in~\cite{shen2016integrated,yao2019optimization} for thrust optimization, and an economic MPC (EMPC) was proposed in~\cite{yang2019energy}. An energy-optimal path-following control strategy was developed in~\cite{yang2020robust}, which computes yaw angle with LOS guidance law for reduced path-following error and optimizes the vehicle surge speed for energy minimization. However, the aforementioned approaches only deal with 2D motion control (i.e., in the horizontal/vertical plane).

Considering the great need to perform 3D path-following maneuvers for AUVs to achieve various underwater missions (e.g., mapping and inspection), in this paper, we extend our previous work in~\cite{yang2020robust} for energy-optimal 3D path-following control under ocean currents. A two-stage controller structure is utilized. In the first stage, the vehicle propulsion energy for surge, heave, and pitch controls is first estimated and then used to compute the optimal setpoints for surge velocity, heave velocity, and pitch angle. The LOS guidance law calculates the yaw angle setpoint. In the second stage, a decoupled control is adopted to track the setpoints, where two MPCs are used to achieve the control in the horizontal and vertical planes. Extensive simulations on different mission profiles and current conditions are performed to verify the proposed approach.




\section{Model and Problem Formulation} \label{section.2}



\subsection{Simulation Model}
An earth-fixed frame $\{ e \}$ and a body-fixed frame $\{ b \}$ are used to describe the AUV motion under ocean currents. Defined in $\{ e \}$, the positions and orientations of an AUV is denoted as $\text{\boldmath$\eta$} = [x, y, z, \phi, \theta, \psi]^T \in \mathbb{R}^{6}$. The velocities (relative to the ground) ($\text{\boldmath$\nu$} = [u, v, w, p, q, r]^T \in \mathbb{R}^{6}$), and control inputs ($\text{\boldmath$\tau$} \in \mathbb{R}^{6}$) are represented in $\{ b \}$. 

The following assumptions are made for modeling the effect of currents on vehicle motions: i) the currents are constant and irrotational, ii) the current velocity component in the $z$ direction is negligible, and iii) the magnitude of currents is smaller than the maximum vehicle speed (relative to water). Note that above assumptions do not hold in general but can be satisfied in a deep-sea environment~\cite{liu2005patterns}, which is the designated operating environment for our testbed AUV, DROP-Sphere developed for deep ocean benthic optical mapping~\cite{iscar2018towards} . Based on the assumptions, the current velocities in $\{ e \}$ and the current velocities in $\{ b \}$ are then given as $\text{\boldmath$V$}_c = [V_c^x, V_c^y, 0, 0, 0, 0]^T \in \mathbb{R}^{6}$ and  $\text{\boldmath$\nu$}_c = [u_c, v_c, w_c, 0, 0, 0]^T \in \mathbb{R}^{6}$, respectively. $\text{\boldmath$V$}_c = \textbf{J}(\text{\boldmath$\eta$}) \text{\boldmath$\nu$}_c $, where $\textbf{J}(\text{\boldmath$\eta$}) \in \mathbb{R}^{6\times6}$ is the coordinate transformation matrix. The vehicle velocities relative to the current (i.e., relative velocities) are defined as $\text{\boldmath$\nu$}_r \triangleq \text{\boldmath$\nu$} - \text{\boldmath$\nu$}_c  = [u_r, v_r, w_r, p, q, r]^T \in \mathbb{R}^{6}$. 

The dynamics and kinematics of AUVs in 3D environments under currents are given by~\cite{caharija2016integral}
\begin{subequations} \label{eq:Vehicle Simulation Mdoel} 
\begin{equation}  \label{eq:Vehicle Kinematics} 
\dot{\text{\boldmath$\eta$}} = \textbf{J}(\text{\boldmath$\eta$})\text{\boldmath$\nu$} _r+ \textbf{V}_c,
\end{equation}
\begin{equation}  \label{eq:Vehicle Dynamics} 
\textbf{m}_t\dot{\text{\boldmath$\nu$}_r}+\textbf{f}_c(\text{\boldmath$\nu$}_r)+\textbf{f}_h\text{\boldmath$\nu$}_r+\textbf{f}_g(\text{\boldmath$\eta$}) = \text{\boldmath$\tau$},
\end{equation}  
\end{subequations}
where $\textbf{m}_t\in \mathbb{R}^{6\times6}$, $\textbf{f}_c(\text{\boldmath$\nu$}_r) \in \mathbb{R}^{6},\textbf{f}_h \in \mathbb{R}^{6\times6}$, and $\textbf{f}_g(\text{\boldmath$\eta$}) \in \mathbb{R}^{6}$ represent the vehicle total mass, Coriolis and centripetal force, hydrodynamic damping force, and hydrostatic force matrices, respectively. See~\cite{prestero2001verification} for the detailed expressions. Only the linear and diagonal components are considered in added mass and hydrodynamic damping matrices. 

Based on the thruster allocation of DROP-Sphere, the control inputs are related to the thrusts by
\begin{equation} \label{eq:Control Input Vector}
\text{\boldmath$\tau$} =\left[                 
T^1 + T^2,   
     0, 
    T^3 + T^4, 
    0,
    (T^3-T^4)l_1,
    (T^1-T^2)l_2
\right]^T,
\end{equation}
where $T^1$ and $T^2$ are the horizontal thrusters at port and starboard, $T^3$ and $T^4$ are the vertical thrusters at fore and aft, $l_1$ is the distance between vertical thrusters and the midship, and $l_2$ is the distance between horizontal thrusters and the center line. We assume that propulsion energy accounts for the most energy use during vehicle operation. The following relationship is adopted to model the power consumption~\cite{yang2021active}:
\begin{equation} \label{eq:power conversion}
h_p(T^i) = \alpha(T^i)^2,\; \text{for }i = 1,2,3,4,
\end{equation}
where $\alpha$ is the power conversion ratio.


\subsection{Problem Formulation}
 
Consider that an AUV is operating in a 3D obstacle-free environment subject to currents. The paths to be followed is defined by the straight-lines connecting successive waypoints in a given set of waypoints $\mathcal{WP}\triangleq\{\text{WP}_i \in \mathbb{R}^3 ,\; i = 0,...,N_w \}$, where $\text{WP}_i = (\text{WP}^x_i,\text{WP}^y_i,\text{WP}^z_i)$ contains the $x$, $y$, and $z$ locations of a waypoint, and $N_w$ is the total number of waypoints. $\text{WP}_0$ corresponds to the initial vehicle location. The objective of the 3D energy-optimal path-following control of AUVs is to find the sequences of thrusts $\{ T^i\}$ that (i) ensures the path following in 3D spaces and (ii) minimize the total propulsion energy. In this study, we assume that there will be negligible error in the knowledge of vehicle states and ocean current velocities at the vehicle location, which can be achieved using advanced sensor fusion algorithms and appropriate sensors~\cite{hegrenaes2011model}.


\section{3D energy-optimal path following control} \label{section.3}


\subsection{Controller Architecture} \label{section.3A}

\begin{figure}[!ht]
\centering
\includegraphics[width=3.4in]{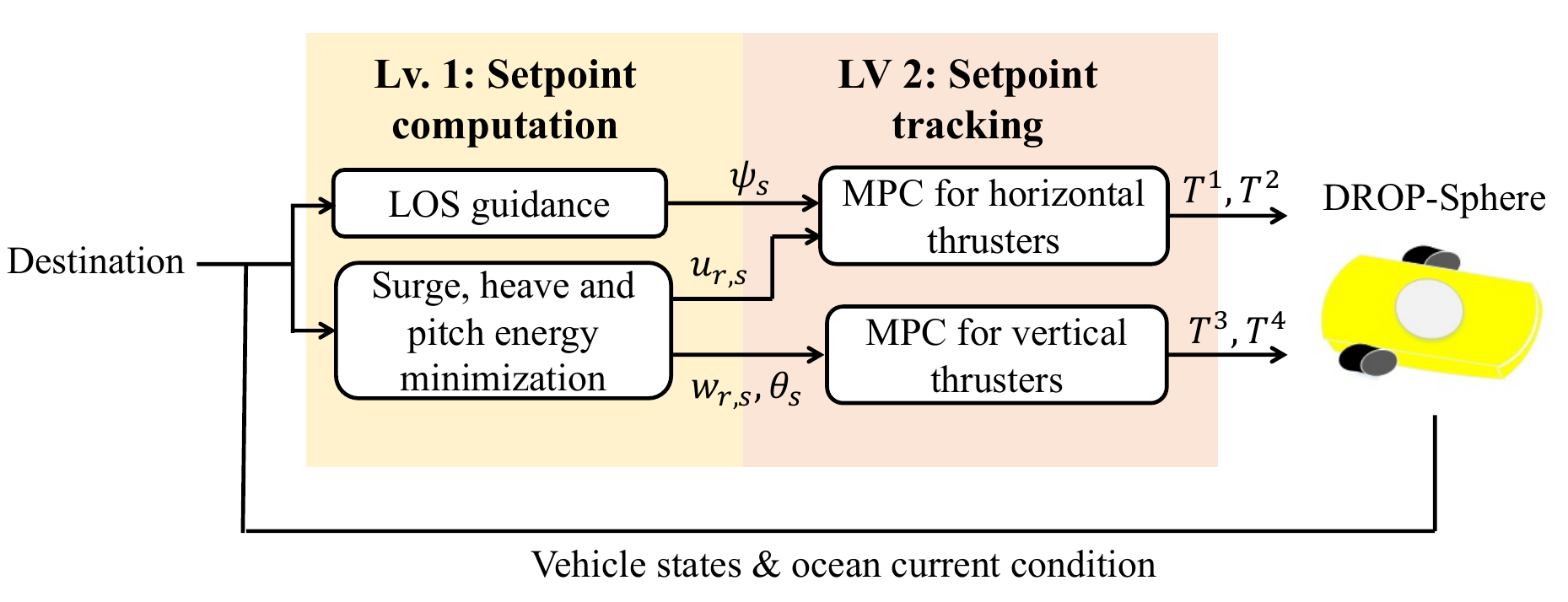} \vspace{-0.2cm}
\caption{Overall controller structure of the proposed energy-optimal control} 
\label{fig:control diagram} 
\end{figure}

Following the energy-optimal path-following control proposed in~\cite{yang2020robust}, a two-stage controller structure shown in Fig.~\ref{fig:control diagram} is adopted. In the first stage, based on the vehicle operating condition and the paths to follow, the desired setpoints are calculated to minimize the vehicle energy and reduce the path-following error. In the second stage, given the desired setpoints, MPCs compute the thruster inputs that minimize the setpoint tracking error. By separating the thrust computation into two stages, the modularized controller structure will require lower computation, making the approach real-time feasible on resource-limited AUV platforms~\cite{gao2015hierarchical,guo2018computationally,ravi2021two}. Meanwhile, the unmodeled dynamics neglected in the modularized design may lead to a sub-optimal solution, as we analyzed in~\cite{yang2018real}. \vspace{-0.2cm}

%
 \vspace{-0.5cm}
%


\subsection{Setpoint computation} \label{section.3B}

For DROP-Sphere thruster allocation, four setpoints, i.e., relative surge velocity setpoint ($u_{r,s}$), relative heave velocity setpoint ($w_{r,s}$), pitch angle setpoint ($\theta_s$), and yaw angle setpoint ($\psi_s$), are required to guide the 3D motion of the vehicle. In this study, we decompose the setpoint computation into i) the yaw angle setpoint computation with LOS and ii) surge velocity, heave velocity, and pitch angle setpoint computation using optimization for the reasons as follows.
\begin{itemize}
    \item \textbf{Natural separation of the transient and steady-state dynamics}: The energy for yaw control is spent only in transients, while the energy for surge, heave, and pitch controls is persistently consumed. Since minimal variations in the vehicle heading will be required during most mission time~\cite{caharija2016integral}, the energy-saving potentials of optimizing the yaw motion is limited.
    \item \textbf{Computation load reduction}: To optimize the yaw motion, yaw dynamics must be considered, leading to increased computation and reduced real-time feasibility.
    \item \textbf{Handling the underactuation}: Given that the vehicle is underactuated in the horizontal plane (i.e., no control for sway motion), designing control for path following is inherently challenging. The decomposition facilitates the use of some well-established approaches, e.g., LOS guidance~\cite{lekkas2013line}, to handle the underactuation by properly choosing the yaw angle setpoint.
\end{itemize}


%
\begin{figure*}[!ht]
\centering
\includegraphics[width=6.6in]{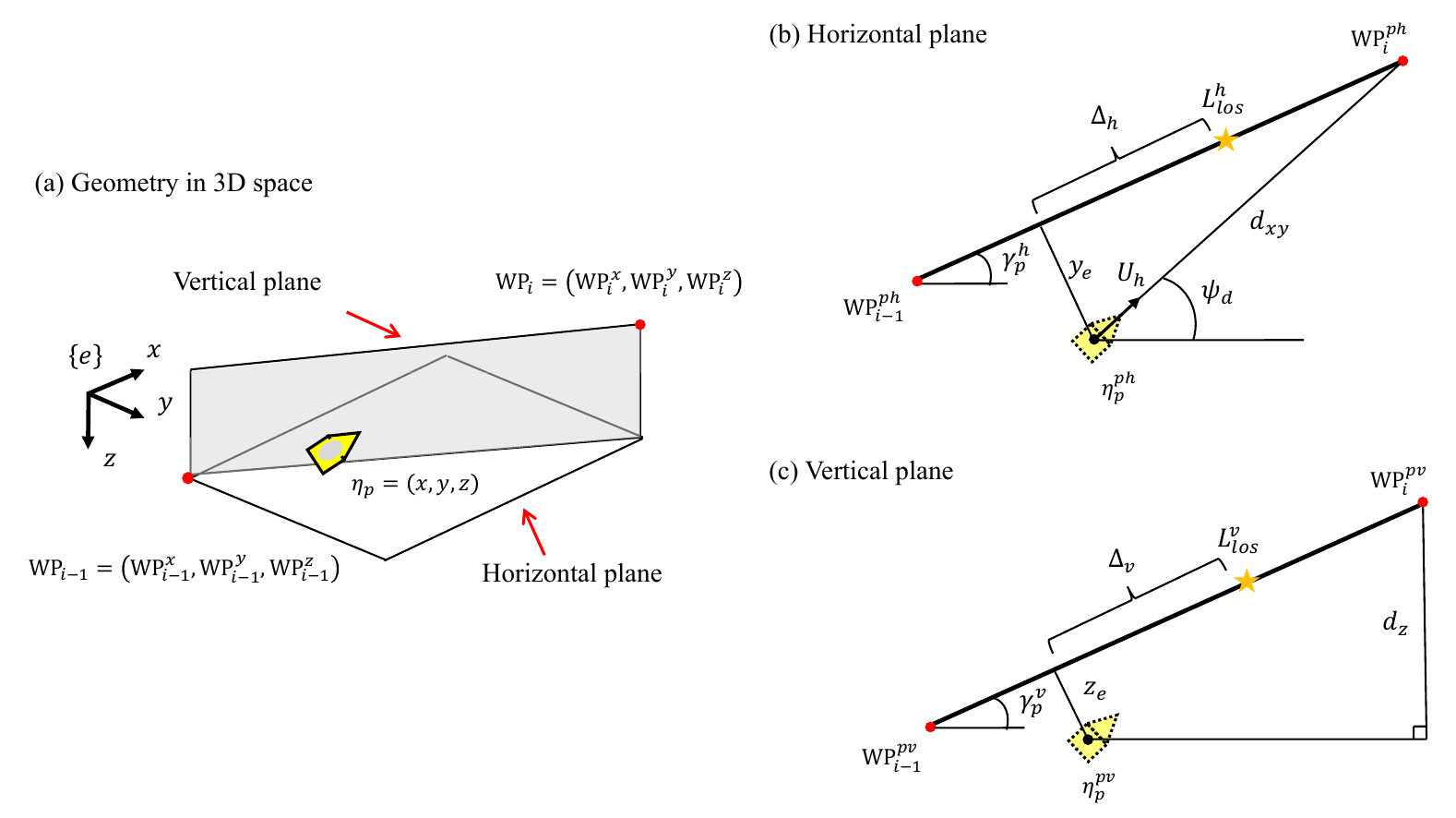} \vspace{-0.2cm}
\caption{Illustration of the horizontal plane, the vertical plane, and the LOS guidance. $(\cdot)^{ph}$ and $(\cdot)^{pv}$ denotes the projection of a (3D) point onto the horizontal and vertical plane, respectively.} 
\label{fig:Illustration of the geometric relationships under different settings} 
 \vspace{-0.5cm}
\end{figure*} 
%


\subsubsection{\textbf{Yaw angle setpoint computation}} The geometry of the LOS guidance in the horizontal plane is illustrated in Fig~\ref{fig:Illustration of the geometric relationships under different settings}(b). Mimicking the behaviors of a hemlsman, the LOS guidance drives a vessel towards an LOS position to ensure the path convergence~\cite{oh2010path}.  Given a path between the waypoints to visit and the last-visited waypoint, the LOS position in the horizontal plane ($L^h_{los}$) is defined as a point located on the path at a lookahead distance ($\Delta_h$) ahead of the vehicle's projection onto the path. The LOS guidance law then computes the yaw angle setpoints under currents as follows~\cite{lekkas2013line}.
\begin{equation} \label{eq:LOS-MPC yaw angle}
\psi_{s}=\gamma^h_{p}-\tan^{-1}(y_e/\Delta_h)-\beta,
\end{equation}
where $\gamma^h_{p} = \si{atan2}(\text{WP}^y_{i}-\text{WP}^y_{i-1},\text{WP}^x_{i}-\text{WP}^x_{i-1})$ is the path-tangential angle in the horizontal plane, the function \si{atan2} returns an arc tangent angle from $-\pi$ to $\pi$, the cross-track error $y_e = -(x-\text{WP}^x_{i-1})\sin\gamma^h_p + (y-\text{WP}^y_{i-1})\cos\gamma^h_p$ is the distance from the vehicle location to the path, and $\beta = \si{atan2}(v_r+v_c,(u_r+u_c)\cos\theta+(w_r+w_c)\sin\theta)$ is the sideslip angle under currents.


\subsubsection{\textbf{Relative surge velocity, relative heave velocity, and pitch angle setpoint optimization}} To optimize the relative surge velocity, relative heave velocity, and pitch angle setpoints, a parameterization of the vehicle energy will be required. The following assumptions are made to derive a model for approximating energy consumed:
1) The roll and sway motion is negligible, i.e., $\dot{p} = p = \phi = \dot{v}_r = v_r = 0$; 2) The surge, heave, pitch, and yaw dynamics are in steady-state, i.e., constant $u_r$, $w_r$, $\theta$, and $\psi$. The second assumption generally does not hold for arbitrary 3D motion, but it holds approximately in deep-sea survey or mapping missions, as discussed early in Section~\ref{section.3B}.


Based on the above assumptions, the simplified vehicle model is given as
\begin{subequations} \label{eq:Model for energy minimization}
\begin{equation} 
T^1+T^2 = (W-B)\sin \theta + X_uu_r,
\end{equation} 
\begin{equation} 
T^3+T^4 = (W-B)\cos \theta + Z_ww_r,
\end{equation}
\begin{equation} 
(T^4-T^3)l_2 = (z_bB-z_gW)\sin \theta,
\end{equation}
\begin{equation} 
\dot{z} = -\sin \theta u_r + \cos \theta w_r,
\end{equation}
\end{subequations}
%
where $W$ is the vehicle weight, and $B$ is the vehicle buoyancy. For DROP-Sphere, $W < B$, i.e., the vehicle is positive buoyant. $X_u$ and $Z_w$ are the surge and heave drag coefficients, respectively. $z_b$ and $z_g$ are the $z$ locations of the buoyancy center and gravity center, respectively. Since the yaw motion is in steady-state, i.e., $T^1 = T^2$, the thruster inputs can be approximated by solving \eqref{eq:Model for energy minimization} as
\begin{subequations} \label{eq:approximated thruster inputs}
\begin{equation} \label{eq:approximated horizontal thruster inputs}
T_a^1 = T_a^2 = \frac{1}{2}((W-B)\sin \theta + X_uu_r),
\end{equation}  
\begin{equation} 
\begin{split} \label{eq:approximated vertical thruster inputs}
T_a^{3,4} = \frac{1}{2}  ((W-B) &\cos \theta + Z_ww_r) \\ 
 &\mp \frac{1}{2l_2} ((z_bB-z_gW)\sin \theta). 
\end{split}
\end{equation} 
\end{subequations}
Using \eqref{eq:power conversion} and \eqref{eq:approximated thruster inputs}, the power consumption of all thrusters is then computed as
\begin{equation} \label{eq:approximated poweer consumption}
P^t = \sum^{4}_{i=1}h_p(T_a^i). 
\end{equation}

For energy consumption computation, the travel time is also required, which can be approximated using the vehicle speed towards the desired waypoint~\cite{yang2020robust}. Denote $U_c$ and $\psi_c$ as the magnitude and direction of the ocean current in the horizontal plane. $\psi_d$ is the direction of the line between the vehicle current location and the desired waypoint (see Fig.~\ref{fig:Illustration of the geometric relationships under different settings}(b)). The velocity component of ocean currents along and perpendicular to the line between the vehicle location and the desired waypoint are $V^a_c = U_c\cos\psi_{cd}$ and $V^p_c = U_c\sin\psi_{cd}$, respectively, where $\psi_{cd} = \psi_c - \psi_d$. The vehicle speed towards the LOS position under current then is derived as
\begin{equation}  \label{eq:vehicle horizontal speed to the destination}
U_h = \sqrt{V_{uw}^2-(V^p_c)^2}+V^a_c,
\end{equation}
where $V_{uw} = \cos\theta u_r+\sin \theta w_r$ is the vehicle relative velocity along its heading in the horizontal plane. Combining \eqref{eq:approximated poweer consumption} and \eqref{eq:vehicle horizontal speed to the destination}, the vehicle energy use is approximated by 
\begin{equation} \label{eq:energy approximation}
E(u_r,w_r,\theta,U_c,\psi_{cd}) = \frac{P^t \cdot d_{xy}}{U_h}, 
\end{equation}
where $d_{xy} = \sqrt{(\text{WP}^y-y)^2+(\text{WP}^x-x)^2}$ is the distance to the desired waypoint in the horizontal plane. Since the approximation in \eqref{eq:energy approximation} considers only the energy in the steady state, the yaw control energy is not captured in \eqref{eq:energy approximation}.

The optimal relative surge velocity, relative heave velocity, and pitch angle setpoints is then computed as follows:
\begin{subequations} \label{eq:surge, heave and pitch angle optimization}
\begin{equation} 
u_{r,s}, w_{r,s}, \theta_s = \mathop{\argmin}_{u_r,w_r,\theta}  E(u_r,w_r,\theta,U_c,\psi_{cd}),
\end{equation}
subject to 
\begin{equation} \label{eq:vertical translation constraint}
\frac{d_{xy}}{U_h} = \frac{d_z}{\dot{z}},
\end{equation}
\begin{equation} \label{eq:vehicle speed constraint}
U_h > 0,
\end{equation}
\end{subequations}
where $d_z = \text{WP}^z-z$. The inequality constraint \eqref{eq:vehicle speed constraint} prevents the vehicle relative velocity along its heading from being too small so that the vehicle travel speed will always be well-defined and positive under currents. The equality constraint \eqref{eq:vertical translation constraint} is meant to ensure path following. Given that constraints can not be directly imposed on the vehicle's future positions in the above optimization, we request the vehicle to take the same time to reach the desired waypoint in the horizontal and the vertical plane. Note that we allow a large path following error in~\eqref{eq:surge, heave and pitch angle optimization} when the vehicle is far from the desired waypoint in the horizontal plane as avoiding drastic maneuvers after a course transition can save energy. 



\subsection{MPC for setpoint tracking} \label{section.3C}
Considering that the coupling between the vehicle dynamics in the horizontal and vertical planes can be weak~\cite{hong2013online}, a decoupled control scheme is employed for the setpoint tracking to reduce the computation. The MPC for horizontal thrusters computes the left and right thrusts based on the yaw angle and relative surge velocity setpoints. The MPC for vertical thrusters takes in the pitch angle and relative heave velocity setpoints to compute the fore and aft thrusts. The MPC is chosen for two reasons. First, it can incorporate input constraints when computing the thrusts. Moreover, for DROP-Sphere, multiple thrusters are involved in 3D maneuvering. MPC has the flexibility to systematically address the multivariate optimization by formulating a proper cost. 

In particular, the horizontal motion MPC is formulated as
\begin{equation}  \label{eq:horizontal MPC}
\begin{split}
\mathop{\min}_{\{ T^{1}\},\{ T^{2}\}} \sum_{k=1}^{N}(\lambda_h (u_{r,k|t}-u_{r,s})^2+(\psi_{k|t}-\psi_{s})^2), 
\end{split}
\end{equation}
where $(\cdot)_{k|t}$ is the $k$-step ahead prediction made at time instant $t$, $N$ is the length of the prediction horizon, and $\lambda_h$ is a constant factor penalizing the tracking error in the relative surge velocity. Note that since the yaw angle setpoint depends on the surge velocity, $\lambda_h$ is required to be larger than $1$ so that the surge motion will be stabilized first, facilitating a faster convergence to the desired waypoint in the horizontal plane. The above optimization is subject to i) thruster limits and ii) the discrete AUV dynamics in the horizontal plane assuming zero heave, roll, and pitch motion. 

For the vertical motion MPC, it is formulated as
\begin{equation}  \label{eq:vertical MPC}
\begin{split}
\mathop{\min}_{\{ T^{3}\},\{ T^{4}\}} \sum_{k=1}^{N}(\lambda_v (w_{r,k|t}-w_{r,s})^2+(\theta_{k|t}-\theta_{s})^2), 
\end{split}
\end{equation}
where $\lambda_v$ is a constant factor penalizing the tracking error in the relative heave velocity. The constraints for the vertical motion MPC are i) thruster limits and ii) the discrete AUV dynamics in the vertical plane assuming constant surge velocity and zero sway and roll motion.


\section{Performance evaluation} \label{section.4}


\subsection{Benchmark algorithm: 3D LOS-based control} \label{section.4A}

The 3D LOS-based control uses the two-stage controller structure in Fig.~\ref{fig:control diagram}. The yaw angle setpoint is determined using the LOS guidance law presented in \eqref{eq:LOS-MPC yaw angle}, and the relative heave velocity setpoint is set to zero. The following LOS guidance law in the vertical plane is used to compute the pitch angle setpoint~\cite{lekkas2013line} (see Fig.~\ref{fig:Illustration of the geometric relationships under different settings}(c) for an illustration):
\begin{equation} \label{eq:LOS-MPC pitch angle}
\theta_{s}=\gamma^v_{p}+\tan^{-1}(z_e/\Delta_v)+\alpha,
\end{equation}
where $\gamma^v_{p} = \si{atan2}(\text{WP}^z_{i-1}-\text{WP}^z_{i},((\text{WP}^x_{i}-\text{WP}^x_{i-1})^2+(\text{WP}^y_{i}-\text{WP}^y_{i-1})^2)^{1/2})$ is the path tangential angle in the vertical plane, $z_e = ((x-\text{WP}^x_{i-1})^2+(y-\text{WP}^y_{i-1})^2)^{1/2}\sin\gamma^v_p + (z-\text{WP}^z_{i-1})\cos\gamma^v_p$ is the cross-track error in the vertical plane, $\Delta_v$ is the look-ahead distance in the vertical plane, and $\alpha= \si{atan2}(w_r+w_c,u_r+u_c)$ is the attack angle under currents. Based on the relative heave velocity setpoint (i.e., zero) and pitch angle setpoint in \eqref{eq:LOS-MPC pitch angle}, the relative surge velocity setpoint is optimized by
\begin{equation}  \label{eq:surge optimization}
u_{r,s} = \mathop{\argmin}_{u_r}  E(u_r,w_{r,s},\theta_s,U_c,\psi_{cd}),
\end{equation}
subject to $U_{h} > 0$. The 3D LOS-based control represents a conventional approach achieving 3D motion control, and only the surge velocity is optimized to reduce the vehicle energy consumption. The setpoint tracking of the 3D LOS-based control is the same as that proposed in Section~\ref{section.3C}.


\subsection{Performance comparison} \label{section.4B}

Two mission profiles are considered to compare the control approaches: 
\begin{enumerate}
    \item \textbf{Lawnmower-type mission}: The $x$, $y$, and $z$ locations of the waypoints are $\text{WP}_{i}\in \{(0,0,0)$, $(30,0,10)$, $(30,10,7)$, $(0,10,1)$, $(0,20,2)$, $(30,20,6)$, $(30,30,5)$, $(0,30,3) \} ~m$ to represent a mapping mission near the sea bottom.
    \item \textbf{Inspection mission}: The $x$, $y$, and $z$ locations of the waypoints ${\text{WP}_i}, i = 0, ..., 12$ in the inspection mission are
    \begin{equation} \label{eq:Circular waypoints}
    \begin{dcases*}
        \text{WP}_i^x = (7\cos\frac{\pi(2i+ 9)}{6} + 7) ~m,\\
        \text{WP}_i^y = (7\sin\frac{\pi(2i+ 9)}{6} + 7) ~m, \\
        \text{WP}_i^z = (2i) ~m.
    \end{dcases*}
    \end{equation}
    This setup represents an inspection mission around an oil platform, e.g., spar platforms.
\end{enumerate}
The vehicle is initialized with zero velocities and heading towards $\text{WP}_1$. The simulation is conducted in MATLAB/Simulink. A course transition will be performed once the distance between the vehicle and the target waypoint is less than $2~\si{m}$. The prediction horizon $N = 10$ steps, and the time step size $\Delta t =  0.1~\si{s}$. $\lambda_h$ and $\lambda_v$ are both chosen as $50$. The lookahead distance $\Delta_h = \Delta_v = 2~\si{m}$.

\begin{table*}
\begin{spacing}{1.0}
\centering 
\caption{Performance comparison with the 3D LOS-based approach (the 3D cross-track error is computed by averaging the cross-track error in 3D spaces when the vehicle is away from the waypoint by $2~\si{m}$)} 
\label{table:Performance comparison with the 3D LOS-based approach}
\begin{tabular}{@{}c|c|ccc|ccc@{}} 
\toprule
\multirow{2}{*}{Current condition} & \multirow{3}{*}{Method} & \multicolumn{3}{c|}{\tabincell{c}{Lawnmower-type mission}} & \multicolumn{3}{c}{\tabincell{c}{Inspection mission}}  \\ \cline{3-8} 
   \multirow{1}{*}{[$V_c^x$, $V_c^y$] $(\si{m/s})$}  & & Energy & \tabincell{c}{Travel time} & \tabincell{c}{3D Cross \\ -track error }  & Energy & \tabincell{c}{Travel time }  & \tabincell{c}{3D Cross \\ -track error } \\ \hline
\multirow{2}{*}{[0.0417, 0.0963]} & Proposed &  $2483.1~\si{J}$ &  $1029.5~\si{s}$ &  $0.1468~\si{m}$ & $2093.8~\si{J}$ & $575.3~\si{s}$ & $0.2008~\si{m}$ \\
 & LOS &  $3053.7~\si{J}$ &  $839.9~\si{s}$ &  $0.0782~\si{m}$ & $2811.4~\si{J}$ & $383.1~\si{s}$ & $0.1250~\si{m}$\\ \hline 
 \multirow{2}{*}{[0.0841, -0.1718]} & Proposed &  $3946.7~\si{J}$ &  $704.9~\si{s}$ &  $0.1438~\si{m}$ & $2886.8~\si{J}$ & $391.4~\si{s}$ & $0.2341~\si{m}$ \\
 & LOS &  $4313.4~\si{J}$ &  $619.8~\si{s}$ &  $0.1038~\si{m}$ & $3444.8~\si{J}$ & $294.1~\si{s}$ & $0.1601~\si{m}$ \\ \hline 
 \multirow{2}{*}{[-0.0342, -0.0678]} & Proposed &  $2437.6~\si{J}$ &  $1209.8~\si{s}$ &  $0.0978~\si{m}$ & $1976.9~\si{J}$ & $673.0~\si{s}$ & $0.1817~\si{m}$ \\
 & LOS &  $3041.6~\si{J}$ &  $963.0~\si{s}$ &  $0.0650~\si{m}$ & $2722.5~\si{J}$ & $417.1~\si{s}$ & $0.1279~\si{m}$ \\ \hline 
 \multirow{2}{*}{[-0.0541, 0.1382]} & Proposed &  $3119.3~\si{J}$ &  $848.2~\si{s}$ &  $0.1534~\si{m}$ & $2532.6~\si{J}$ & $462.9~\si{s}$ & $0.2323~\si{m}$ \\
 & LOS &  $3500.1~\si{J}$ &  $704.5~\si{s}$ &  $0.1193~\si{m}$ & $3053.8~\si{J}$ & $332.8~\si{s}$ & $0.1383~\si{m}$ \\ \hline 
\multirow{2}{*}{[0, 0]} & Proposed &  $1982.1~\si{J}$ &  $1490.8~\si{s}$ &  $0.1112~\si{m}$ & $1798.2~\si{J}$ & $834.3~\si{s}$ & $0.1600~\si{m}$ \\ 
 & LOS &  $2634.4~\si{J}$ &  $1108.9~\si{s}$ &  $0.0430~\si{m}$ & $2488.8~\si{J}$ & $461.5~\si{s}$ & $0.1343~\si{m}$  \\ \bottomrule
\end{tabular} 
\end{spacing} 
 \vspace{-0.5cm}
\end{table*} 

\begin{figure}[!h]
\centering
\includegraphics[width=3.3in]{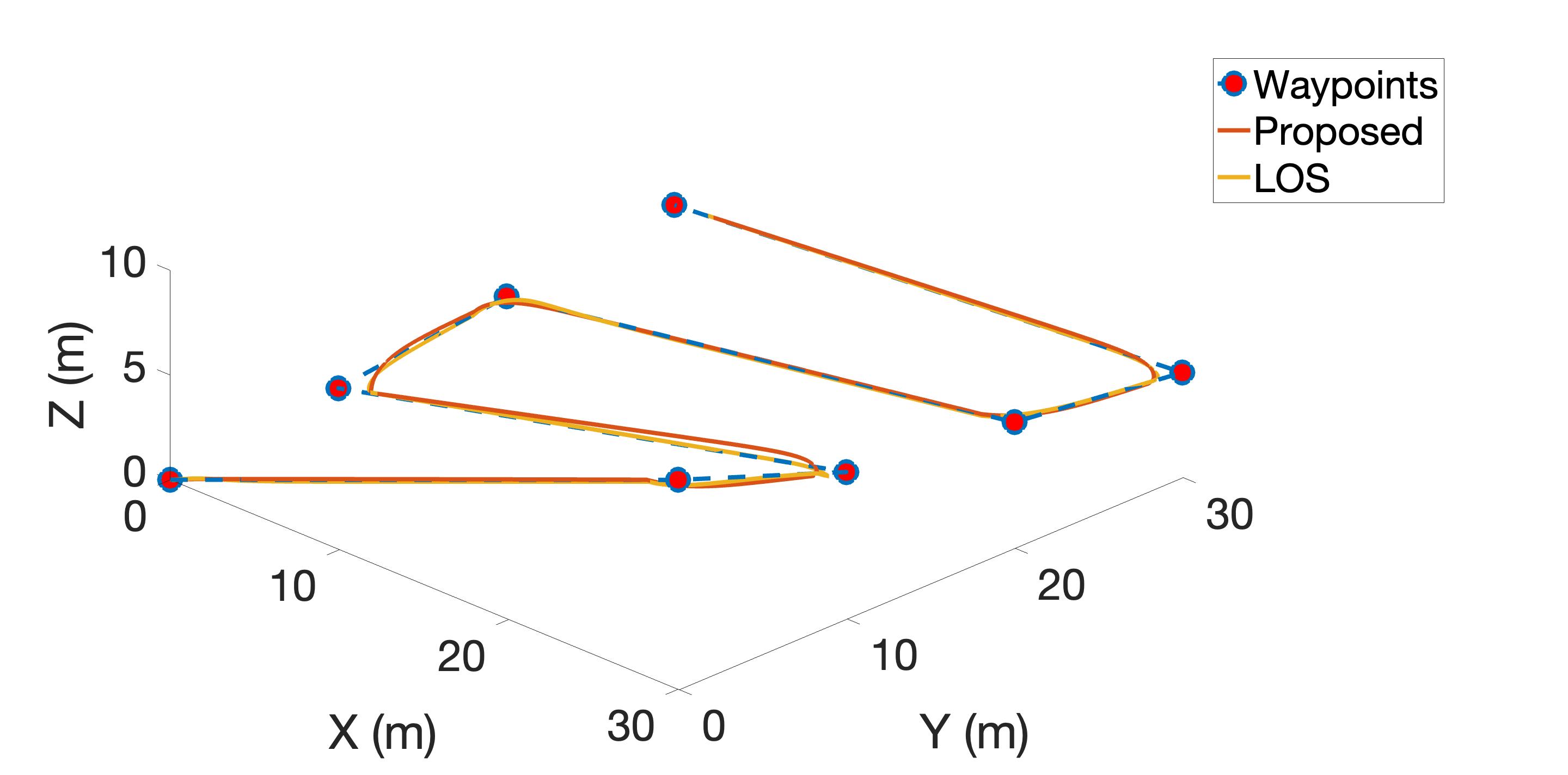} \vspace{-0.1cm}
\caption{Vehicle trajectory (lawnmower-type mission $[V_c^x, V_c^y]$ = $[0.0417,0.0963]~\si{m/s}$)} 
\label{fig:Vehicle trajectory (lawnmower-type mission)} 
\end{figure}

\begin{figure}[!h]
\centering
\includegraphics[width=3.3in]{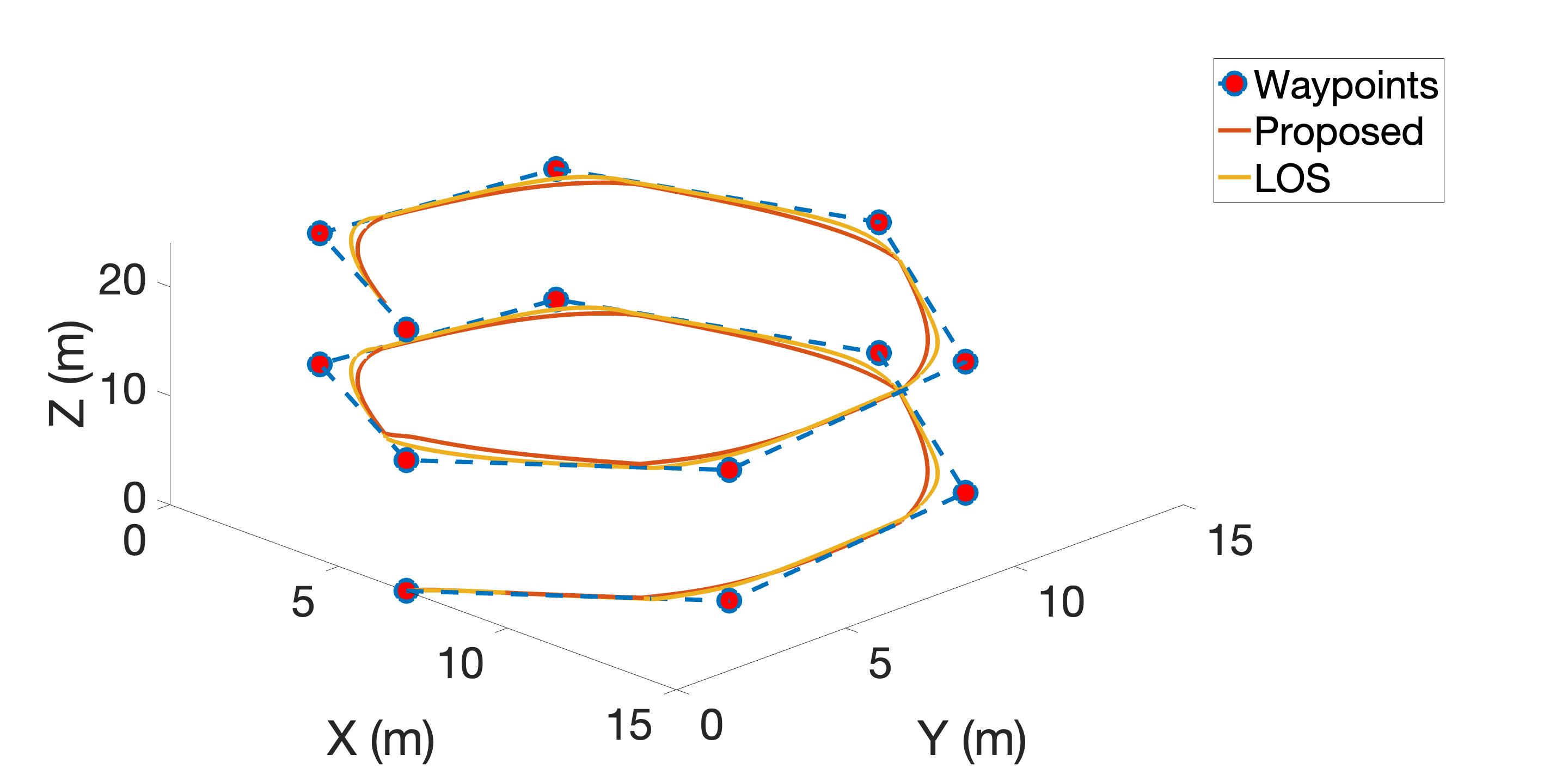} \vspace{-0.1cm}
\caption{Vehicle trajectory (inspection mission $[V_c^x, V_c^y]$ = $[0.0417,0.0963]~\si{m/s}$)}  \vspace{-0.7cm}
\label{fig:Vehicle trajectory (spar inspection mission)} 
\end{figure}

The performance of two control strategies under five different flow conditions is summarized in Table~\ref{table:Performance comparison with the 3D LOS-based approach}. Note that since the scale of both missions is small, the current velocities are set as constant in all simulations. The vehicle trajectories in the lawnmower-type and inspection mission for the case where $[V_c^x, V_c^y] = [0.0417,0.0963]~\si{m/s}$ are given in Fig.~\ref{fig:Vehicle trajectory (lawnmower-type mission)} and Fig.~\ref{fig:Vehicle trajectory (spar inspection mission)}, respectively. It can be seen that both approaches can sequentially visit all the waypoints and follow the desired paths. The proposed approach reduces vehicle propulsion energy compared to the LOS-based approach on both mission profiles under all current conditions. An average of $13.81 \%$ reduction in energy consumption is achieved. On the downside, the travel time of the proposed approach can be higher as the proposed approach only aims at minimizing energy consumption. In addition, since the proposed approach does not enforce a small depth tracking error until it is close to the desired waypoint in the horizontal plane, the depth tracking error can be large when the vehicle is far from the desired waypoint (see Fig.~\ref{fig:Vehicle trajectory (lawnmower-type mission)} and Fig.~\ref{fig:Vehicle trajectory (spar inspection mission)}), leading to a larger 3D cross-track error. \vspace{-0.3cm}
%

%
 \vspace{-0.2cm}

\begin{figure}[!ht]
\centering
\includegraphics[width=3.3in]{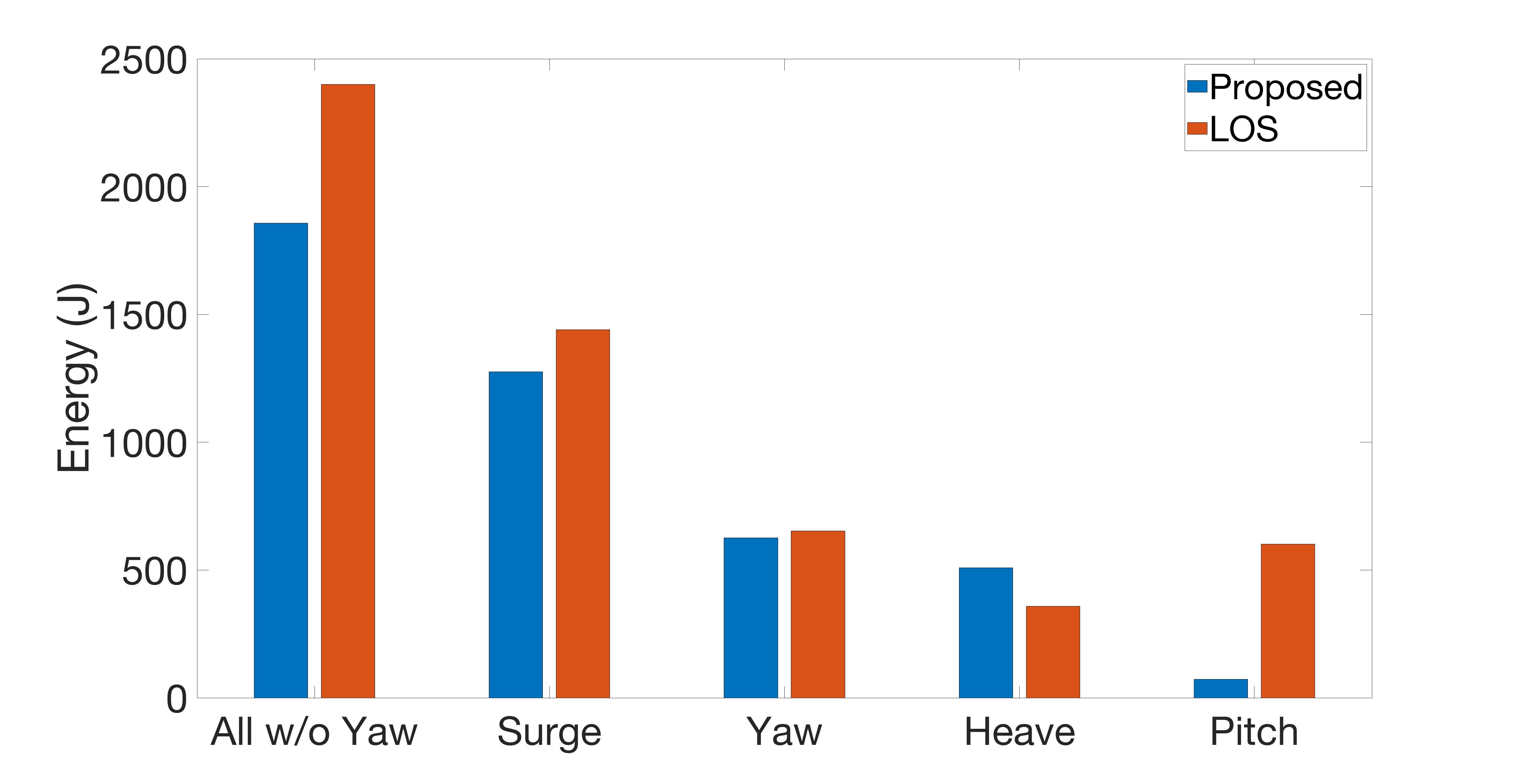}  \vspace{-0.1cm}
\caption{Energy analysis (lawnmower-type mission $[V_c^x, V_c^y]$ = $[0.0417,0.0963]~\si{m/s}$)} 
\label{fig:Energy analysis (lawnmower-type mission)} 
\end{figure}

To better understand the characteristics of an energy-efficient maneuver, the energy used to control the motion in different DOFs for the solutions in Fig.~\ref{fig:Vehicle trajectory (lawnmower-type mission)} is also provided in Fig.~\ref{fig:Energy analysis (lawnmower-type mission)}, which leads to the following observations:
\begin{itemize}
    \item The proposed energy approximation can capture the major trade-off in propulsion energy during a path following mission, leading to reduced energy use for surge, heave, and pitch controls.
    \item The proposed approach can not ensure energy reduction for yaw control as the energy approximation in \eqref{eq:energy approximation} does not consider the yaw energy consumption.
    \item Employing non-zero relative heave velocities results in more energy-efficient maneuvers. Compared to the 3D LOS-based control, the proposed approach reduces energy by consuming more heave energy and less surge and pitch energy, which implies a larger heave speed, a lower surge speed, and a smaller pitch angle.
\end{itemize} \vspace{-0.4cm}
 \vspace{-0.1cm}


\section{Conclusions and Future Work} \label{section.5}
In this paper, an energy-optimal path-following approach is proposed to control the motion of autonomous underwater vehicles in 3D spaces with ocean currents. The proposed approach computes the optimal thrusts in two stages. In the first stage, the relative surge velocity, relative heave velocity, and pitch angle setpoints are optimized by minimizing the approximated energy required to follow the path till the next course transition under ocean currents. Assuming the vehicle dynamics are negligible, the energy approximation is achieved by characterizing the vehicle thrusts and parameterizing the vehicle travel speed. The yaw angle setpoint is computed with the line-of-sight guidance law to ensure a robust path following performance under currents. Given the optimized setpoints, model predictive control (MPC) is adopted in the second stage to track the setpoints. Extensive performance evaluations are conducted on a lawnmower-type mission and a spar inspection mission under different current conditions. It is shown that the proposed approach can achieve substantial energy efficiency improvements, an average $13\%$ saving for the conditions considered. 

In this work, only simulations are performed to verify the proposed approach. Experimental validations will be conducted in the future to further demonstrate the effectiveness of the proposed approach. In addition, the ocean currents and vehicle location are known perfectly in this study, which can be hard to achieve for some low-cost AUV platforms (e.g., DROP-Sphere). Future studies about the effects of uncertainties in the ocean currents and vehicle location on the control performance are also of great interest. \vspace{-0.05cm}


\bibliographystyle{IEEEtran}
\bibliography{Reference}


\end{document}